\documentclass[aip,
rsi,
 amsmath,amssymb,
 reprint%
]{revtex4-1}

\usepackage{graphicx}
\usepackage{dcolumn}
\usepackage{bm}
\usepackage{color}

\begin{document}

\title{Measurement of the angle dependence of magnetostriction in pulsed magnetic fields using a piezoelectric strain gauge}% Force line breaks with \\

\author{Xiaxin Ding}
\email{xding@lanl.gov}
\affiliation{National High Magnetic Field Laboratory (NHMFL), Materials Physics and Applications (MPA)-Magnet (MAG) Group, Los Alamos National Laboratory (LANL), Los Alamos, New Mexico 87545, USA}
\author{Yi-Sheng Chai}
\email{yschai@iphy.ac.cn}
\affiliation{Beijing National Laboratory for Condensed Matter Physics, Institude of Physics, Chinese Academy of Sciences, Bejing 100190, P.R. China}
\author{Fedor Balakirev}
\author{Marcelo Jaime}
\affiliation{National High Magnetic Field Laboratory (NHMFL), Materials Physics and Applications (MPA)-Magnet (MAG) Group, Los Alamos National Laboratory (LANL), Los Alamos, New Mexico 87545, USA}
\author{Hee Taek Yi}
\author{Sang-Wook Cheong}
\affiliation{Rutgers Center for Emergent Materials and Department of Physics and Astronomy, Piscataway, New Jersey 08854, USA}
\author{Young Sun}
\affiliation{Beijing National Laboratory for Condensed Matter Physics, Institude of Physics, Chinese Academy of Sciences, Bejing 100190, P.R. China}
\author{Vivien Zapf}
\affiliation{National High Magnetic Field Laboratory (NHMFL), Materials Physics and Applications (MPA)-Magnet (MAG) Group, Los Alamos National Laboratory (LANL), Los Alamos, New Mexico 87545, USA}

\date{\today}% It is always \today, today,

\begin{abstract}
We present a high resolution method for measuring magnetostriction in millisecond pulsed magnetic fields at cryogenic temperatures with a sensitivity of $1.11\times10^{-11}/\sqrt{\rm Hz}$. The sample is bonded to a thin piezoelectric plate, such that when the sample's length changes, it strains the piezoelectric and induces a voltage change. This method is more sensitive than a fiber-Bragg grating method. It measures two axes simultaneously instead of one. The gauge is small and versatile, functioning in DC and millisecond pulsed magnetic fields. We demonstrate its use by measuring the magnetostriction of Ca$_3$Co$_{1.03}$Mn$_{0.97}$O$_6$ single crystals in pulsed magnetic fields. By comparing our data to new and previously published results from a fiber-Bragg grating magnetostriction setup, we confirm that this method detects magnetostriction effects. We also demonstrate the small size and versatility of this technique by measuring angle dependence with respect to the applied magnetic field in a rotator probe in 65 T millisecond pulsed magnetic fields.

\end{abstract}

%\pacs{Valid PACS appear here}
\maketitle

%\tableofcontents

\section{INTRODUCTION}

Most magnetic and electronic materials exhibit a measurable magnetostriction. Magnetostriction, $e.g.$ dilatometry in magnetic fields, is a change in lattice dimensions in response to a magnetic field ($H$) that lowers the magnetic and electronic energy at the expense of the energy of deforming the lattice~\cite{overview}. For example, magnetostriction can modify the Fermi surface, the exchange interactions, or local spin-lattice couplings. Magnetostriction is thus an important thermodynamic quantity as well as a powerful tool for understanding magnetic behavior of materials and spin-lattice coupling. Magnetostriction is one of the more sensitive techniques for detecting and tracking magnetic-field-induced transitions particularly in pulsed magnetic field. Fig. 1(a) shows a schematic diagram of the magnetostriction effect. The magnetic material placed in a magnetic field along the 1-axis undergoes a structural distortion, which causes a small change of in length $\Delta L_1$, yielding a strain $\lambda_{11} = \Delta L_1/L_1$ where $L_1$ is the original length. Additionally, there are also strain changes perpendicular to the magnetic field direction, $\lambda_{21} = \Delta L_2/L_2$ and $\lambda_{31} = \Delta L_3/L_3$. The strains $\lambda_{11}$, $\lambda_{21}$ and $\lambda_{31}$ are the magnetostriction. The signed ratios of transverse to longitudinal strains $-\lambda_{21}/\lambda_{11}$ and $-\lambda_{31}/\lambda_{11}$ (Poisson's ratio) are usually less than 0.5, indicating that the change along the field direction is usually larger than that along the transverse directions.

Multi-shot pulsed magnets are designed to produce magnetic fields up to 100 T with pulse lengths ranging from 0.01 to 1 s, allowing researchers to study interesting physics phenomena under ultrahigh magnetic fields~\cite{doan}. Measuring magnetostriction in pulsed magnetic field requires overcoming the challenge of measuring on short timescales (65 T short pulsed magnets at the NHMFL have 10 ms rise times), electrical and mechanical noise caused by the rapidly changing field, and eddy currents in metallic materials. Existing methods for measuring magnetostriction in pulsed magnetic fields include:

(1) Resistive foil strain gauges. This strain gauge consists of meandering wire attached to an insulating flexible film. The gauge is glued to the sample. The resistance of the meandering wire in the gauge is sensitive to strain. Though this technique is insensitive to vibrations, its resolution is limited and its strong magnetoresistance must be well calibrated. A resolution of 5$\times$10$^{-6}$ per reading in pulsed magnetic fields has been reported by Algarabel $et$ $al$.~\cite{strain}. 

(2) Capacitance dilatometry. This is a widely used technique for high sensitivity data acquisition in DC magnetic fields~\cite{Schmiedeshoff}. The length change of the sample is obtained by measuring the change in capacitance between a variable plate attached to one end of the sample, and a fixed plate. However in pulsed magnetic fields, the technique achieves low sensitivity due to its vibration-sensitivity, the constrained space inside pulsed magnets, eddy currents in the capacitor plates, the cell's own dilatometry background when the cell is not at a uniform well-controlled temperature, and the motion of bubbles in the helium past the capacitor plates in applied magnetic fields. $\lambda$ as small as $\sim10^{-5}$ per reading can be resolved using capacitance dilatometry at 10-20 kHz in pulsed fields, or $\sim10^{-7} /\sqrt{{\rm Hz}}$~\cite{capacitance}. 

(3) Atomic force microscope (AFM) piezocantilever dilatometry. AFM piezocantilevers are commonly used in torque measurement in pulsed magnetic fields~\cite{torque}. Park $et$ $al$. first reported this technique for magnetostriction measurements~\cite{piezocantilever}. The tip of the cantilever rests on the sample and the change in length is reflected in the change in resistance of the cantilever as it is deformed by sample strain. This device is ultracompact and the size of the lever arm is just 0.4 $\times$ 0.05 $\times$ 0.005 mm$^3$. However, the cantilevers are fragile and difficult to mount on the sample. Moreover, the device is sensitive to the vibrational noise and the cantilevers can be broken by thermal contractions of the sample with respect to the mounting device. This technique also have a cell background. The resolution is very high, however, the exact number is not known.

(4) Optical fiber strain gauges or Fiber Bragg gratings (FBG). This recently-developed technique has opened the way for high resolution optical-based magnetostriction measurements in pulsed magnetic fields, capable of resolving strains on the order of 10$^{-7}$ with a full bandwidth of 47 kHz, or $\sim10^{-9} /\sqrt{{\rm Hz}}$~\cite{fiber, sensor}. In this method the sample is glued to an optical fiber that has a Bragg grating (equally spaced lines) etched into it. Laser light Bragg-diffracts off the grating, providing a sensitive and intrinsically calibrated measure of the change in grating spacing and thus sample strain. This technique delivers data that is only minimally affected by electromagnetic noise and mechanical vibrations. Without any metal parts, there is no eddy current heating. One limitation of the FBG method, however, is that the fiber has a minimum bending radius, thus this technique can't measure transverse magnetostriction in small bore pulsed magnets, and cannot measure a continuous angle dependence of the magnetostriction. Another drawback is the need to glue the sample to the fiber. The glue can fail or absorb part of the magnetostriction.

(5) Piezoelectric transducers. This is the method reported here. Its basic principle of operations is to bond the sample to a thin piezoelectric material that senses the change in length of the sample via a change in its ferroelectric polarization, $e.g.$ its voltage. In 1992, Levitin $et$ $al$. reported a version of this method using quartz as the piezoelectric in pulsed magnetic fields. It was not widely disseminated and predated modern piezoelectrics and the current intense interest in pulsed-field magnetostriction measurements~\cite{Russia}. This paper suggests that the $\Delta L/L$ resolution can achieve 10$^{-9}$ per reading.

Other probes of magnetostriction include measurements of magneto-optical Kerr effect in thin films upon applying strain~\cite{MOKE}, which is an indirect probe of magnetostriction, and also X-ray~\cite{XrayA, XrayJ} and neutron diffraction. In particular Larmor neutron diffraction has recently emerged as a technique that can detect lattice parameter changes with 10$^{-6}$ precision~\cite{neutron}. Magnetostriction is not only a probe of fundamental physics of materials, but it is also attractive for a number of applications including transducers/motors, torque sensing, as components of multiferroics and other multifunctional devices, and for energy harvesting~\cite{overview,Nan,science,appl,design}.

In this paper, we introduce the implementation of the piezoelectric strain gauge (PSG) method using modern ultra-sensitive piezoelectric materials in pulsed magnetic fields at cryogenic temperatures. The PSG method doesn't need a special sample preparation, satisfies the needs of cost-effectiveness, easy operation, being self-powered, and high resolution. The small foot print is suitable for use in limited space, such as sample rotators in pulsed fields. We discuss the principle of operations and demonstrate its application for measuring the angle dependence of the magnetostriction of Ca$_3$Co$_{1.03}$Mn$_{0.97}$O$_6$ single crystals. It is capable of resolving changes in strain of 1.11$\times10^{-11} /\sqrt{Hz}$. 

\section{PRINCIPLE OF OPERATIONS}

\begin{figure}[t]
\includegraphics[width=8.5cm]{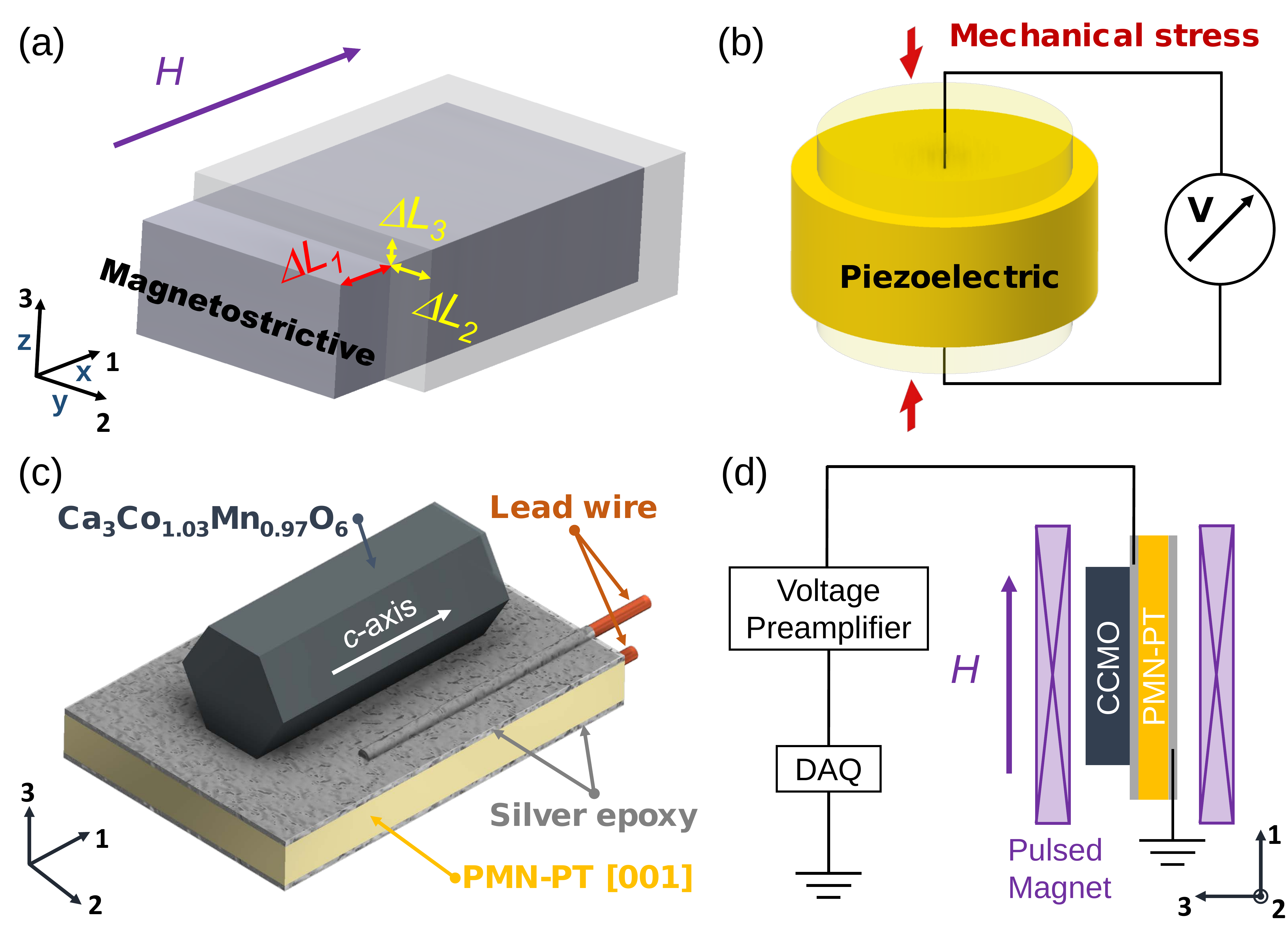}
\caption{\label{fig:fig1}(color online) Schematic illustration of the PSG method. (a) Direct magnetostrictive effect: change $\Delta L$ in the length $L$ of the magnetostrictive sample in response to the magnetic field $H$. For convenience, directions x, y and z are represented by the subscripts 1, 2 and 3, respectively. (b) Direct piezoelectric effect: change in surface charge generated on the opposite surfaces of the piezoelectric material due to a mechanical stress, measured by the voltmeter. (c) ME composite of the magnetostrictive sample CCMO, the piezoelectric material PMN-PT [001] and silver epoxy layers. (d) PSG setup: the ME composite, a pulsed magnet, a low noise voltage preamplifier and a data acquisition system.}
\end{figure}

Similar to magnetostrictive materials, piezoelectric materials have the ability to convert energy between mechanical and electrical forms and are widely used in both experiments and applications. As illustrated in Fig. 1(b), when a mechanical stress is applied to the piezoelectric material, the resulting mechanical deformation generates a change in the surface charge and the voltage across the sample. It is well known that coupling between magnetic or electric orders leads to a magnetoelectric (ME) effect, which is defined as the change of polarization (magnetization) of a material in an external magnetic (electric) field~\cite{multiferroics}. Apart from single-phase ME materials, one way to accomplish the strong ME coupling is an interfacial-coupled laminate composed of the magnetostrictive and piezoelectric layers~\cite{composite}. 

In principle, the strain-mediated ME laminate can directly generate an electrical signal in response to the magnetostrictive effect of the magnetic layer. Accordingly, by bonding a magnetostrictive sample to a piezoelectric material, we are able to develop the PSG method which characterizes the $H$ dependent strain of the magnetic sample by measuring the induced voltage variation of the piezoelectric layer. Its basic principle can be roughly explained as follows. In this case, the piezoelectric material has 2D isotropic properties ($s^p_{11} = s^p_{22}, s^p_{12} = s^p_{21}, d^p_{31} = d^p_{32}$) while the magnetic material is 2D anisotropic ($s^m_{11} \neq s^m_{22}, q^m_{11} \neq q^m_{22} > q^m_{12} = q^m_{21}$). Here $s$ is the elastic compliance, $e.g.$ strain per unit stress, or the reciprocal of Young's modulus. $d$ are the piezoelectric coefficients, $e.g.$ surface charge or electric polarization generated per unit stress. $q$ are the piezomagnetic coefficients $i.e.$, $q^m_{ij} = \delta \lambda_{ij}/\delta H_j$ ($i$,$j$ = 1 and 2). The superscripts $m$ and $p$ represent the magnetostrictive and piezoelectric phase, respectively. The transverse ME coefficient $\alpha_{E,31} = \delta E_3/\delta H_1$ referring to the transverse electric field $E_3$ generated along the 3-axis by an applied $H \parallel$ 1-axis, can be calculated via the following relation~\cite{Srinivasan}:
\begin{eqnarray}
\nonumber \alpha_{E,31} &=& -\frac{kfd^p_{31}(q^m_{11}+q^m_{21})}{\underline{s_{11}}\varepsilon^{T,p}_{33}-2kf(d^p_{31})^2}\\
&\approx& -\frac{kfd^p_{31}(q^m_{11}+q^m_{21})}{\underline{s_{11}}\varepsilon^{T,p}_{33}}
\end{eqnarray}
where $\underline{s_{11}} = f(s^p_{11}+s^p_{12})+k(1-f)(s^m_{11}+s^m_{12})$. $f$ is the volume fraction of the magnetic phase, $f = v^m/(v^p+v^m)$, $v^m$, $v^p$ denote the volume of magnetostrictive and piezoelectric phases, respectively. $k$ is the interface coupling parameter (the coupling factor $k$ =1 for an ideal interface and 0 for the case without friction). $\varepsilon^{T,p}_{33}$ is the permittivity of the piezoelectric phase. Then in the open circuit condition, by sweeping the magnetic field from zero to $H_1$, the accumulated voltage $V_3$ across the piezoelectric layer with a thickness $t$ will be:
\begin{eqnarray}
\nonumber V_3(H_1) &=& \int_{0}^{H_1} t \alpha_{E,31}\delta H_1\\ 
\nonumber &\approx& -\frac{tkfd^p_{31}}{\underline{s_{11}}\varepsilon^{T,p}_{33}}\int_{0}^{H_1} (q^m_{11}+q^m_{21})\delta H_1\\
&=& -\frac{tkfd^p_{31}}{\underline{s_{11}}\varepsilon^{T,p}_{33}}[\lambda_{11}(H_1)+\lambda_{21}(H_1)]
\end{eqnarray}
Similarly,
\begin{eqnarray}
V_3(H_2) \approx -\frac{tkfd^p_{31}}{\underline{s_{22}}\varepsilon^{T,p}_{33}}[\lambda_{22}(H_1)+\lambda_{12}(H_1)]
\end{eqnarray}
where $\underline{s_{22}} = f(s^p_{22}+s^p_{21})+k(1-f)(s^m_{22}+s^m_{21})$. From above equations, we prove that the voltage output $V_3$ is directly proportional to the sum of in-plane strains. According to Poisson's ratio, the $\lambda_{ii}$ term will be dominant in the PSG method. We note that equations (2) and (3) neglect in-plane shear strain. This method doesn't measure the absolute value, and needs to be calibrated.

\section{EXPERIMENTAL DESIGN}

The target compound is Ca$_3$Co$_{1.03}$Mn$_{0.97}$O$_6$ (CCMO). It has a rhombohedral structure composed of alternating Co$^{2+}$ and Mn$^{4+}$ ions in oxygen cages along $c$-axis chains. These chains in turn form a hexagonal lattice in the $ab$ plane~\cite{sample}. Below 15 K, CCMO shows an $\uparrow\uparrow\downarrow\downarrow$ collinear magnetic structure of the alternating Co$^{2+}$ and Mn$^{4+}$ spins along $c$-axis chains at zero magnetic field. Due to the $\uparrow\uparrow\downarrow\downarrow$ spin configuration, the spatial inversion symmetry is broken and a net electric polarization $P$ is induced, making it a single phase multiferroic. Our previous work suggests that there are several metamagnetic phase transitions with anisotropic transition fields. We previously determined that both $c$-axis Ising Co$^{2+}$ and quasi-isotropic Mn$^{4+}$ magnetic ions most likely have $S= \frac{3}{2}$ in CCMO at all magnetic fields, by studying magnetization, electric polarization, magnetostriction and magnetocaloric effect~\cite{Jaewook}. However, we could only measure for $\Delta L \parallel H \parallel c$ using the FBG method. Hence, CCMO is a good test sample for angle dependent magnetostriction measurements by the PSG method.

\begin{figure}[t]
\includegraphics[width=8cm]{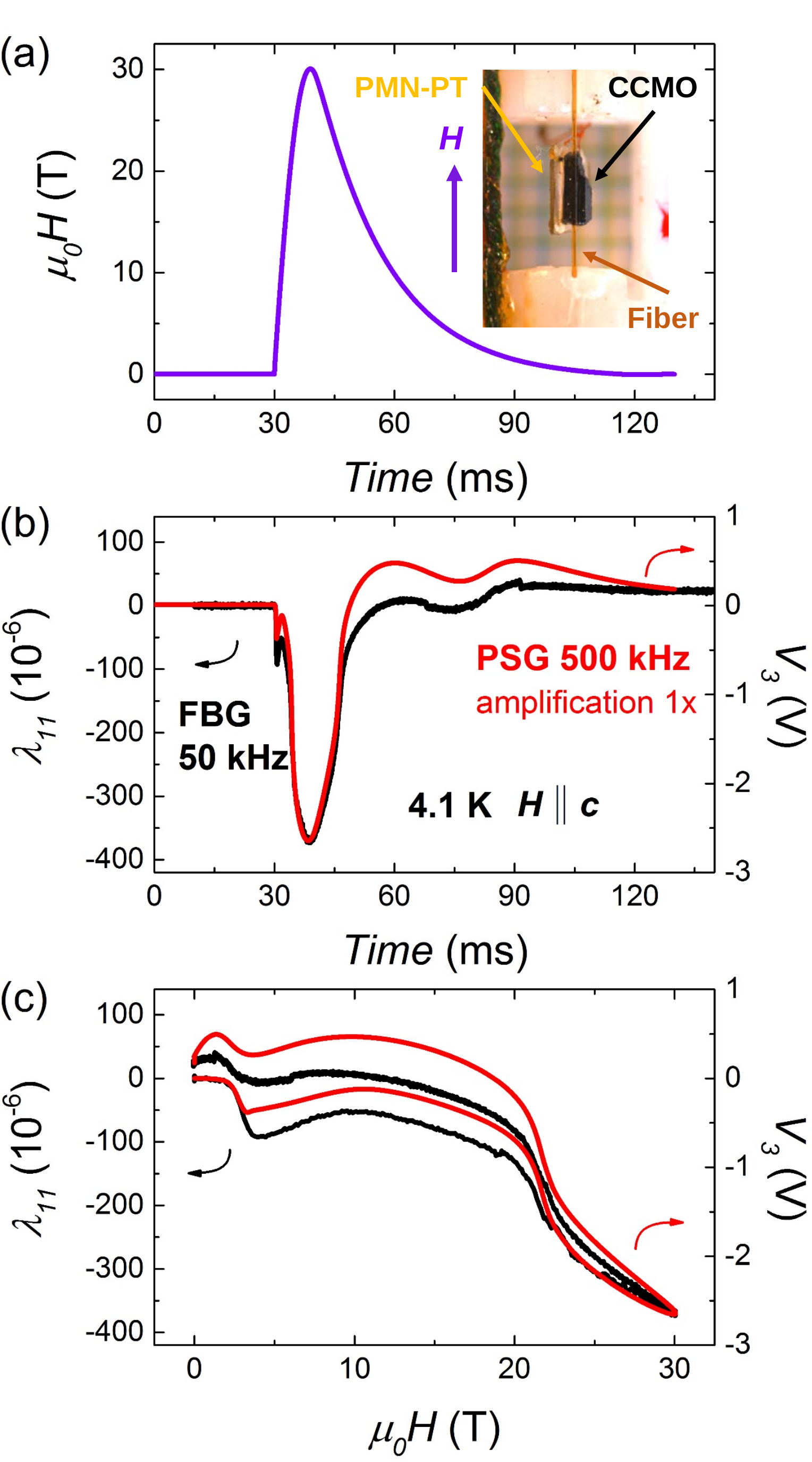}
\caption{\label{fig:fig2}(color online) Simultaneous FBG and PSG measurements of CCMO (a) Time profile of a 30 T field pulse. Inset: single crystal CCMO glued to a 1mm FBG on one side and bonded to a 2$\times$2$\times$0.2 mm$^3$ PSG on the other side. The spacing of the green grid is 1 mm. (c) Time profile of strain $\lambda_{11}$ (black) and ME output voltage $V_3$ (red) measured by FBG and PSG simultaneously at 4.1 K with $H \parallel c$, when exposed to a 30 T pulsed field. The sampling rate of the FBG is 50 kS/s or 50 kHz while that of the PSG is 500 kHz. The amplification of $V_3$ is 1 time. (d) Magnetostriction of CCMO as a function of magnetic field up to 30 T.}
\end{figure}

To prepare the ME composite for measuring CCMO, we chose 0.67 [Pb(Mg$_{1/3}$Nb$_{2/3}$)O$_3$]-0.33 [PbTiO$_3$] (PMN-PT) as the piezoelectric layer, which has the largest piezoelectric coefficient at room temperature. In order to achieve the maximum conversion efficiency, we used [001]-cut PMN-PT single crystals with typical dimensions of 2$\times$2$\times$0.2 mm$^3$ where the 3-axis is along [001] and 1 and 2 are [100] and [010], respectively. This cut and plate geometry gives rise to an isotropic in-plane piezoelectric response (transverse piezoelectric coefficients $d^p_{31} = d^p_{32}$, meaning that the voltage generated along the 3-axis per unit force applied either along the 1- or 2-axis is the same)~\cite{PMNPT}. It also exhibits large electromechanical coupling coefficients, high piezoelectric coefficients, high dielectric constants and low dielectric losses. To achieve effective strain coupling between CCMO and PMN-PT, we polished the contacting surfaces of CCMO and PMN-PT on a 5 micron lapping film. After the polishing process, these two materials were mechanically bonded with sliver conductive epoxy (H20E EPO-TEK), as shown in Fig. 1(c). The sliver epoxy layers also acted as electrodes of PMN-PT. Since the as-bought PMN-PT single crystal had mixed ferroelectric domains, an electric poling treatment along the [001] direction was necessary before cooling the ME composite down to low temperatures. 130 V was applied along the [001] direction for 30 min at room temperatures. 

Then, the ME composite was mounted loosely on a sample probe with two coaxial wires connecting to the sample electrodes. The probe was inserted into the pulsed field magnet bore and cooled down to 1.5 K in $^4$He. A schematic diagram of the sample configuration and the PSG is given in Fig. 1(d). In our design, the ME composite was always mounted on the probe with the magnetic field perpendicular to the electric field (3-axis), so that $H$ always lies in the (1,2) plane or $ca$ plane for CCMO. The angle between $H$ and 1-axis ($c$-axis for CCMO) is defined as $\theta$. With the aid of a SR560 low noise voltage preamplifier (input impedance 100M$\Omega$), we amplified the raw voltage between 1 and 5 times. The millisecond pulsed magnetic field up to 65 T was driven by a capacitor bank at the NHMFL LANL. We performed two sets of measurements in pulsed magnetic fields. In the first set, we used the high resolution 24 bit digitizer (NI PXI-5922) to record the output signal of the PSG, and built a combination probe to measure the magnetostriction of CCMO by FBG and PSG simultaneously. In the second set, we used a standard 12 bit digitizer (NI PCI-5105) to record $V_3$ at different angles.

\section{SIMULTANEOUS FBG AND PSG MEASUREMENTS}

Since the PSG method needs to be calibrated, we built a combination probe that measures the magnetostriction by FBG and PSG simultaneously in pulsed magnetic fields. As shown in the inset of Fig. 2(a), the single crystal CCMO was glued to a 1 mm FBG on one side and a 2$\times$2$\times$0.2 mm$^3$ PSG on another side. The $c$-axis of CCMO was parallel to the field direction (1-axis). To explore the limit of the resolution of the PSG, we used the NI PXI-5922 digitizer with 24 bits at 500 kHz to record the output voltage $V_3$. A typical time profile of a 30 T field pulse is shown in Fig. 2(a). Meanwhile, Fig. 2(b) shows the time dependence of the magnetostriction of CCMO measured by FBG (black) and PSG (red) simultaneously in a 30 T pulsed magnetic field. The sampling rate of the FBG is 50 kHz. $V_3$ was calibrated by the FBG data $\lambda_{11}$. It is clear that $V_3$ responds to the magnetic field instantaneously on the scale of our measurements. The entire time dependent $V_3$ curves are very smooth without spikes, which indicates that our method is not limited by magnetic noise or mechanical vibration generated by the pulsed magnet. Roughly speaking, the time-dependence of $V_3$ is almost the same as the response of $\lambda_{11}$. To verify this, we plot $\lambda_{11}$ and $V_3$ as a function of $H$ in Fig. 2(c). Except for their different magnitude, $\lambda_{11}$ and $V_3$ have very similar $H$-dependences and transition fields. As mentioned above, $V_3$ should be proportional to the magnetostriction of CCMO along the 1-axis, $\lambda_{11} = \Delta L_1/L_1$, however also contains a component of $\lambda_{21}$ (second axis). Therefore, the difference can be attributed to the fact that the FBG measures one axis and the PSG measures two axes of the single crystal.

To show the data quality and resolution of our method, we look into the data more carefully. The rms variation of $\lambda_{11}$ from the FBG at zero field measured over a 20 ms time period is marked by two solid lines in Fig. 3(a). The rms deviation $\sigma_\lambda$ is 1.30 $\times10^{-6}$ per reading, or 2.24$\times10^{-8}/\sqrt{\rm Hz}$ in this system. Fig. 3(c) shows an enlarged view of $V_3$ from the PSG at zero field measured over a 20 ms time period. The rms variation of $V_3$ is $\sigma_V = 5.58 \times10^{-5}$ V, which corresponds to a strain resolution of 7.18 $\times10^{-9}$ per reading, or 1.11 $\times10^{-11} /\sqrt{\rm Hz}$. Thus, we can conclude that the PSG method is more sensitive than the FBG method. Fig. 3(b) and Fig. 3(d) zoom in on the downsweep data of these two techniques near 10 T.

\begin{figure}[!]
\includegraphics[width=8.5cm]{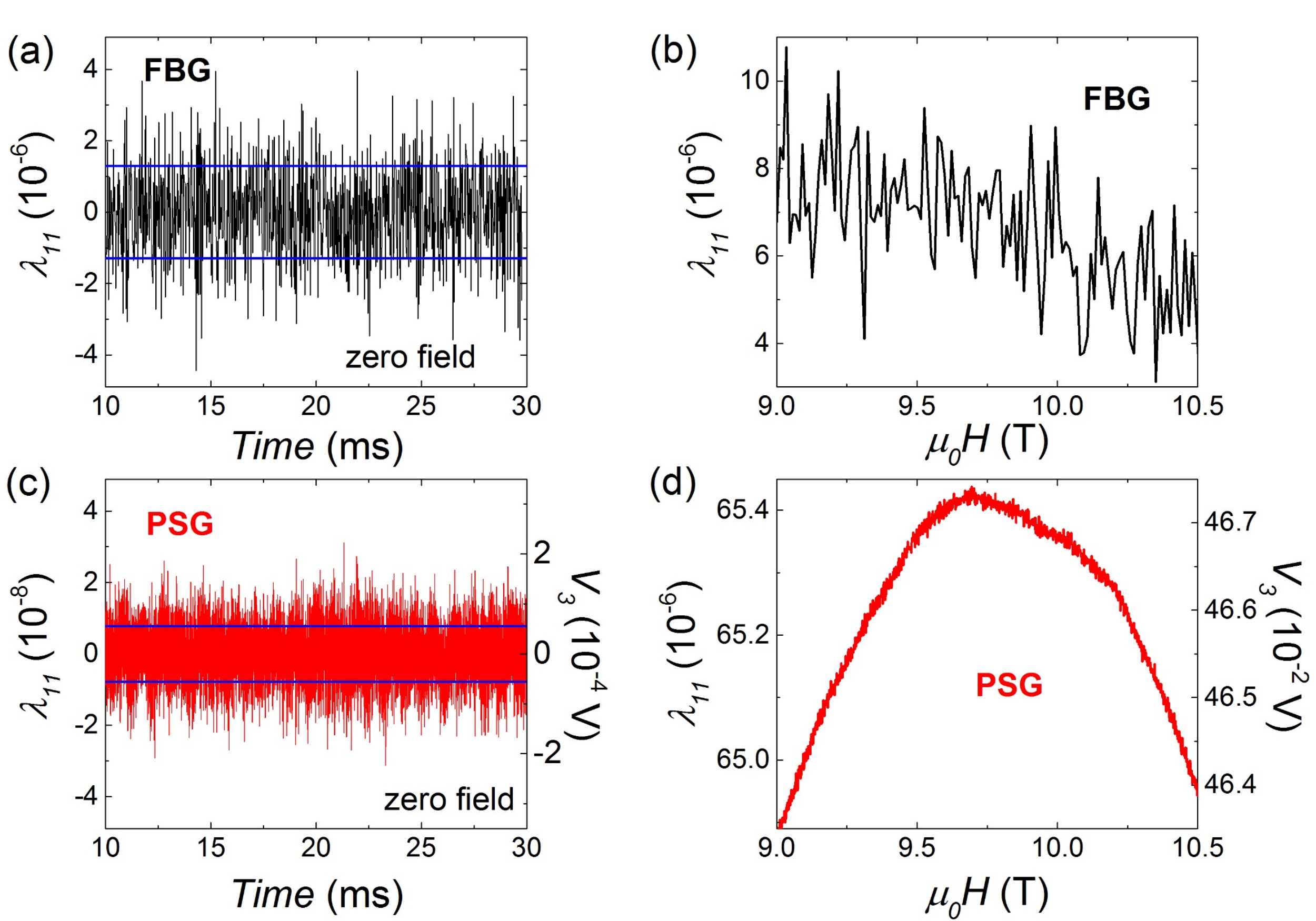}
\caption{\label{fig:fig3}(color online) Comparison of resolution of FBG and PSG: (a) An enlarged view of $\lambda_{11}$ from the FBG at zero field over 20 ms. Two blue horizontal lines mark the rms variation $\sigma_\lambda = 1.30 \times10^{-6}$. (b) An enlarged views of the downsweep data of the FBG near 10 T. (c) An enlarged view of $V_3$ from the PSG at zero field over 20 ms. The rms variation $\sigma_V = 5.58 \times10^{-5}$ V is marked by two blue lines, which corresponds to a strain resolution of 7.18 $\times10^{-9}$. (d)  An enlarged views of the downsweep data of FBG near 10 T.}
\end{figure}

We have investigated the origin of the noise in the PSG. We find that it originates from the SR560 pre-amplifier, since the noise level is the same whether the piezoelectric is disconnected or connected to the preamplifier. Moreover, comparing Fig. 3(c) and (d) we note that the noise is the same whether the magnetic field is pulsing or not. Thus, the noise originates from the preamplifier. The purpose of the preamplifier, even if it is at a gain of 1, is to create better impedance matching between the piezoelectric and the input of the computer's DAQ card. We also note that we filter an 0.25 mV 60 Hz signal and its odd harmonics from all our data using a fast Fourier transform (FFT) analysis.

\section{ANGLE DEPENDENCE OF MAGNETOSTRICTION}

\begin{figure}[!]
\includegraphics[width=8.5cm]{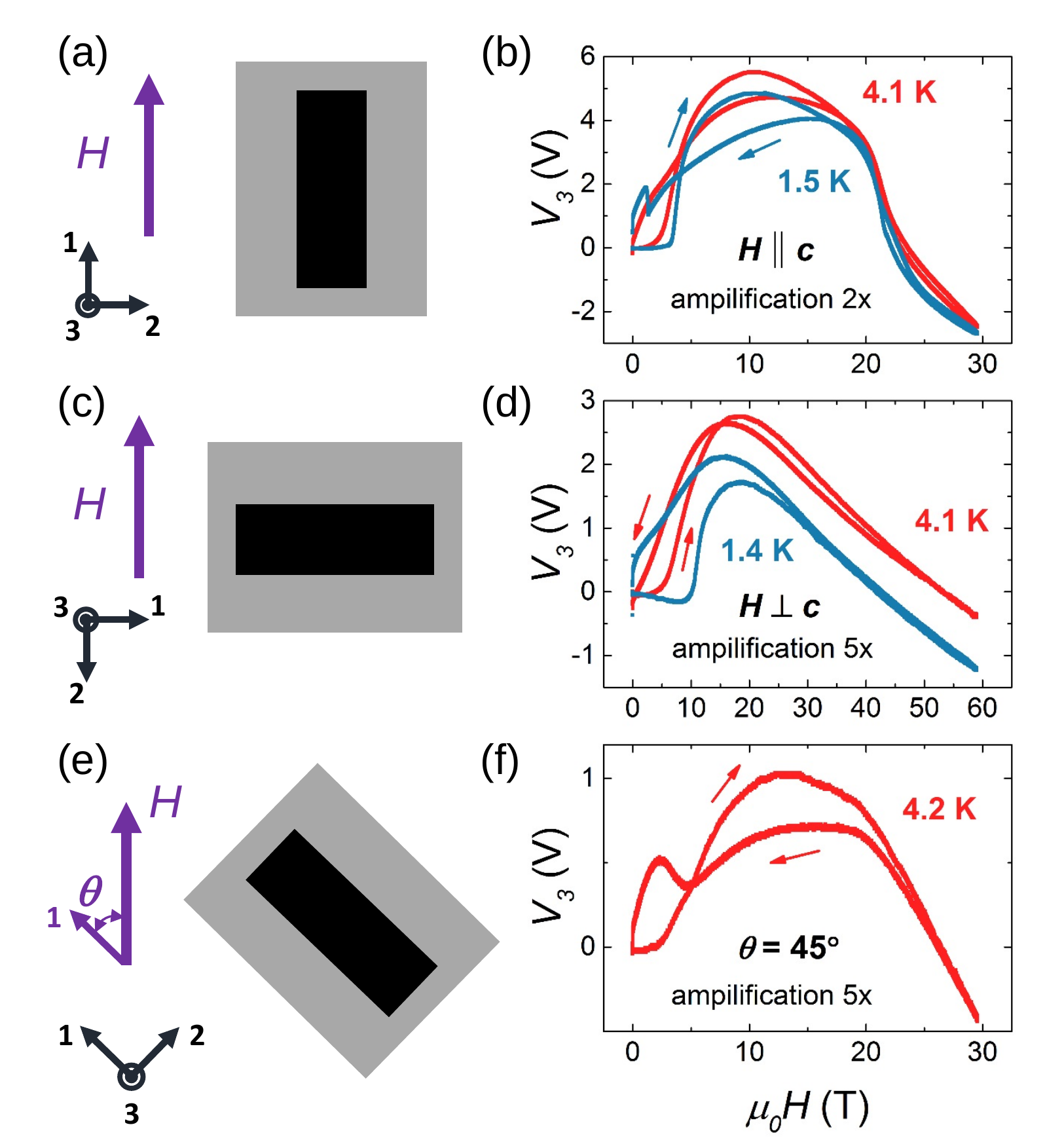}
\caption{\label{fig:fig4}(color online) Schematic diagram of CCMO and PMN-PT in the (1,2) plane with (a) $H \parallel c$, (c) $H \perp c$ and (e) $\theta$ = 45$^{\circ}$. $\theta$ is the angle between $c$-axis and the direction of the magnetic field. The dark shape shows the original shape of CCMO in the $ab$ plane at zero field. The in-plane magnetic field induces strain in CCMO due to the magnetostricive effect, which is mechanically transferred to PMN-PT generating an out-of-plane electrical potential across it. The ME output voltage is plotted as a function of magnetic field with (b) $H \parallel c$ at 1.5 and 4.1 K, (d) $H \perp c$ at 1.4 and 4.1 K, (f) $\theta$ = 45$^{\circ}$ at 4.2 K. The output signal with $H \parallel c$ is amplified 2 times, and the other two are amplified 5 times.}
\end{figure}

In this section we demonstrate that the small footprint and lack of moving parts make the PSG particularly suited for measuring the angle dependence of the magnetostriction in the pulsed magnet. The angle between the $c$-axis of CCMO and the direction of the magnetic field is defined as $\theta$. We used another piece of CCMO from the same batch. Initially, the PSG was parallel to the field direction ($\theta$ = 0$^{\circ}$). Fig. 4(b) shows the field dependence of the PSG signal $V_3$ with $H \parallel c$ at 1.5 and 4.1 K. In the upsweep curve of 1.5 K, the sharp jump up at 5 T and jump down at 20 T are fully consistent with previous reports for the magnetostriction of this material~\cite{Jaewook}. The self-cross feature in the low $H$ region of the 92 T data can be well reproduced by our $V_3$ data. We note that the hysteresis behavior of the $V_3$ curve at low field is different from the published 92 T data due to the difference in magnetic field sweep rate. Our data show a sudden jump at 1.5 T for the downsweep curve. The upsweep and downsweep data match very well above 20 T, which indicates that the self-heating effect must be negligibly small for the PSG method. 

Next, we rotated the entire PSG to $\theta$ = 90$^{\circ}$ and 45$^{\circ}$ positions and measured its $V_3$ as a function of magnetic field, as shown in Fig. 4(c,d) and (e,f), respectively. $\theta$ = 90$^{\circ}$ corresponds to the $H \parallel ab$ in CCMO. Both configurations demonstrate completely different $H$ dependent $V_3$ behaviors from that of the $H \parallel c$ configuration, consistent with the anisotropy expected for Ising-like spins. In particular, for the $\theta$ = 90$^{\circ}$ ($H \perp c$) case at 1.4 K, the upsweep transition field is about 10 T and reaches the peak value at about 20 T, which are consistent with the critical fields observed in the $H \parallel ab$ dependent polarization measurement at 1.5 K~\cite{Jaewook}. For $V_3$ of $\theta$ = 45$^{\circ}$, the magnitude is smaller than the other two configurations. The transition fields for upsweep curve are 5 T and 20 T, which is closer to that of the $H \parallel c$ configuration at 4.2 K

Even though there are many advantages of the PSG method over previous methods, it still has disadvantages. For example, the interface plays an important role in mediating the strain transferred from the sample to the piezoelectric plate. However, the larger piezoelectric plate places more strain on the sample than the very thin fiber of the FBG method. Moreover, both the longitudinal and transverse magnetostrictions appear together in the field dependent behavior. Finally, this method needs to be calibrated for each temperature.

\section{CONCLUSIONS}

We have used a modern ultra-sensitive piezoelectric plate (PMN-PT) as the piezoelectric strain gauge to measure the angle dependence of the magnetostriction of CCMO in pulsed magnetic fields at cryogenic temperatures. There are many advantages to the PSG technique. It is low cost and easy to implement. The lack of moving parts makes it much less sensitive to vibrations than some other methods. We resolved strains of 7.18 $\times10^{-9}$ per reading at 500 kHz, which corresponds to 1.11 $\times10^{-11} /\sqrt{\rm Hz}$, making it the most sensitive method of measuring magnetostriction reported in the literature. The output signal is entirely provided by PMN-PT, leading to a passive sensing without the need for an external power source.

\begin{acknowledgments}

This work is supported by the Laboratory-Directed Research and Development program at Los Alamos National Lab under the auspices of the U.S. Department of Energy (DOE). The National High Magnetic Field Lab Pulsed Field facility is supported by the National Science foundation under cooperative grants DMR-1157490 and DMR-1644779, the U.S. DOE and the State of Florida. Y.-S. Chai and Y. Sun was financially supported by National Natural Science Foundation of China under Grant No. 51725104 and No. 11674384. The work at Rutgers University was supported by the DOE under Grant No. DOE: DE-FG02-07ER46382. We would like to thank Andy Balk for useful discussions.

\end{acknowledgments}

\nocite{*}

\bibliography{PSG}

\end{document}